%% file: CIPANP2018-Leach.tex
\documentclass[12pt]{article}
\usepackage{graphicx}
\usepackage[sort&compress,numbers,square]{natbib}
\usepackage{doi}

\newcommand\pubnumber{CIPANP2018-Leach}
\newcommand\pubdate{\today}

\def\support{\footnote{Work supported by the Department of Energy Office of Science under contract DE-SC0017649.}}
\def\supportc{\footnote{Work supported by the Natural Science and Engineering Research Council of Canada (NSERC).}}

\def\Title#1{\begin{center} {\Large #1 } \end{center}}
\def\Author#1{\begin{center}{ \sc #1} \end{center}}
\def\Address#1{\begin{center}{ \it #1} \end{center}}

\newcommand\pubblock{\rightline{\begin{tabular}{l} \pubnumber\\
         \pubdate  \end{tabular}}}
\newenvironment{Abstract}{\begin{quotation}  }{\end{quotation}}
\newenvironment{Presented}{\begin{quotation} \begin{center} 
             PRESENTED AT\end{center}\bigskip 
      \begin{center}\begin{large}}{\end{large}\end{center} \end{quotation}}
\def\Acknowledgements{\bigskip  \bigskip \begin{center} \begin{large}
             \bf ACKNOWLEDGEMENTS \end{large}\end{center}}

\input econfmacros.tex

\begin{document}
\begin{titlepage}
\pubblock

\vfill
\Title{Weak-Interaction Tests via Precision Superallowed $\beta$-Decay Studies at TRIUMF}
\vfill
\Author{Kyle G. Leach\support}\Address{Department of Physics, Colorado School of Mines, Golden, CO 80401, USA}
\Author{Jason D. Holt\supportc}\Address{TRIUMF, 4004 Wesbrook Mall, Vancouver, BC V6T 2A3, Canada}
\vfill
\begin{Abstract}
Superallowed $\beta$-decay studies provide some of the best constraints on the possibility of additional quark generations, as well as limits on exotic currents in the weak interaction. The three experimental quantities that are required for performing these tests using $0^+\rightarrow0^+$ nuclear decays (branching ratio, half-life, and $Q$-value) can all be measured to high-precision with rare-isotope beams at the TRIUMF-ISAC facility in Vancouver, Canada. This proceeding presents a brief outline of the general experimental techniques used at TRIUMF over the past 15 years, as well as recent theoretical advances towards {\it ab-initio} isospin-symmetry-breaking corrections to superallowed nuclear decays.
\end{Abstract}
\vfill
\begin{Presented}
Conference on the Intersections of Particle and Nuclear Physics\\
Palm Springs, CA, USA, May 29 - June 3, 2018
\end{Presented}
\vfill
\end{titlepage}
\def\thefootnote{\fnsymbol{footnote}}
\setcounter{footnote}{0}

\section{Introduction}
High-precision measurements of nuclear decay properties have proven to be a critical tool in the quest to understand possible physics beyond the Standard Model (BSM)~\cite{Nav13}.  Superallowed $0^+\rightarrow0^+$ nuclear $\beta$-decay data are among the most important to these tests, as they currently provide the most precise determination of the vector coupling strength in the weak interaction, $G_V$~\cite{Har15,PDG16}.  This is possible in this specific decay mode, since the transition operator that connects the initial and final $0^+$ states is independent of any axial-vector contribution to the weak interaction.  In fact, the up-down element of the Cabibbo-Kobayashi-Maskawa (CKM) quark-mixing matrix, $V_{\mathrm{ud}}$, is the most precisely known value in the CKM matrix ($0.021\%$)~\cite{PDG16}, and relies nearly entirely on the high-precision superallowed $\beta$-decay $ft$ values determined through measurements of the half-life, $Q$-value, and branching fraction of the superallowed decay mode~\cite{Har15}.
\newline\indent
In order to obtain the level of precision required for Standard Model tests, corrections to the experimental $ft$-values must also be made to obtain nucleus-independent ${\cal F}t$ values,
\begin{eqnarray}
{\cal F}t\equiv ft(1+\delta_R)(1-\delta_C)=\frac{2\pi^3\hbar^7\ell n(2)}{2G_V^2m_e^5c^4(1+\Delta_R)},
\label{Ft_value}
\end{eqnarray}
where $\delta_R$ is a transition-dependent radiative correction, $\Delta_R$ is a transition-independent radiative correction, and $\delta_C$ is a nucleus-dependent isospin-symmetry-breaking (ISB) correction.  Although relatively small ($\sim1\%$), these corrections are crucial due to the very precise ($\leq0.1\%$) experimental $ft$ values~\cite{Har15}.  The uncertainty on $G_V$, and consequently $V_{ud}$, is presently dominated by the precision of these theoretical corrections, specifically $\Delta_R$ and $\delta_C$.  This proceeding reports on preliminary work towards {\it ab-initio} nuclear-structure calculations of $\delta_C$, as well as a discussion of the experimental work at TRIUMF-ISAC over the past 15 years.

\section{Experimental High-Precision $\beta$-Decay Program}
\begin{figure}[t!]
\centering
\includegraphics[width=\linewidth]{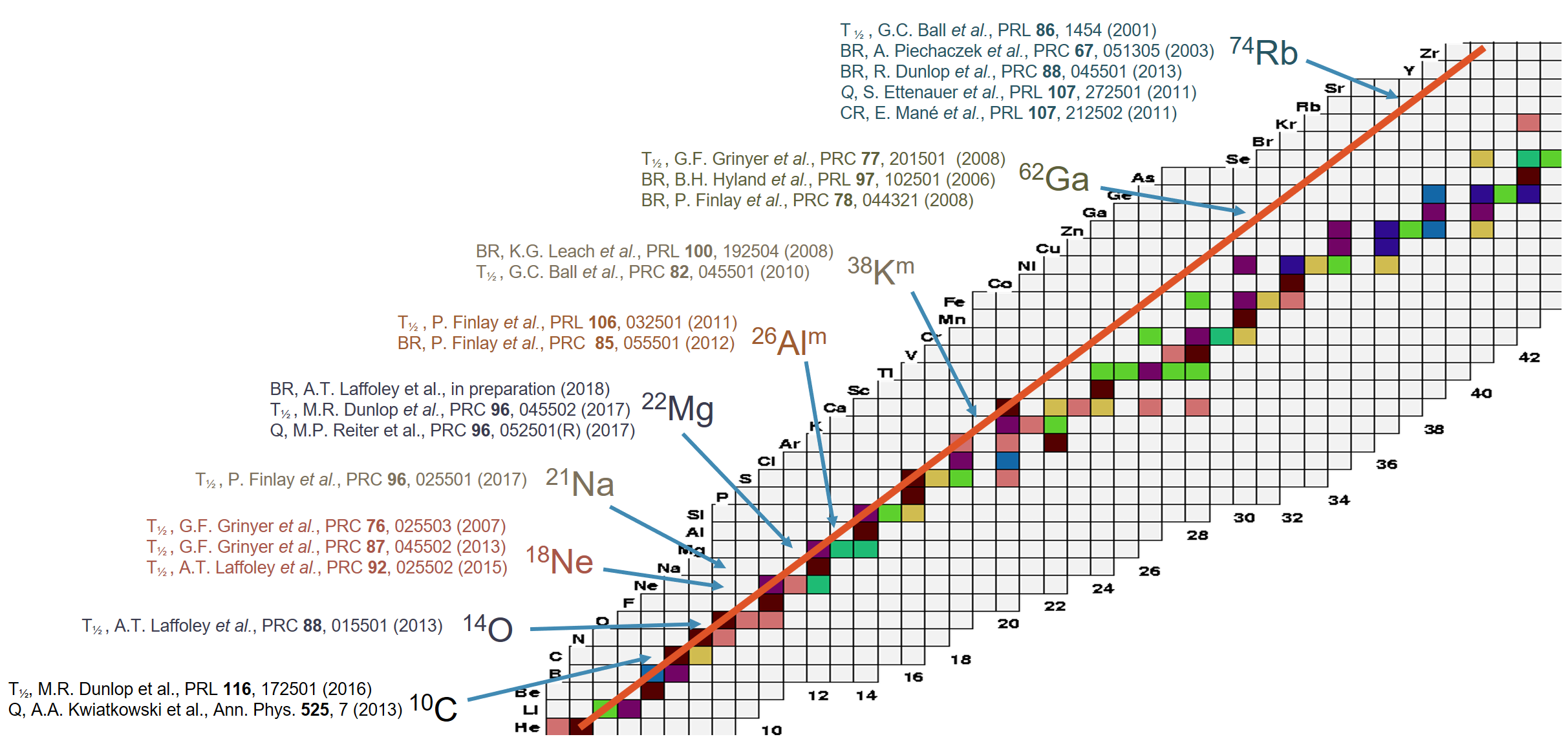}\\
\caption{(Color online) The experimental superallowed decay work performed at TRIUMF-ISAC.  Listed for each nucleus studied are the publications and type of measurement: BR$\rightarrow$ branching ratio, $T_{1/2}\rightarrow$ half-life, and $Q\rightarrow Q$-value.  Figure adapted and updated from Ref.~\cite{Lea11}.}
\label{fig:superallowed}
\end{figure}

The superallowed Fermi decay program has been located at TRIUMF's Isotope Separator and Accelerator (ISAC) facility~\cite{Dil14} in Vancouver, Canada for more than 15 years (Fig.~\ref{fig:superallowed}).  ISAC is an ISOL-type rare-isotope beam (RIB) facility that utilizes a 480~MeV proton driver to induce spallation reactions in a thick production target to generate unstable nuclei.  These nuclei diffuse out of the production target and are subsequently ionized and delivered to the respective experimental apparati at energies between 20-60~keV/$q$.  Critically, this facility is able to produce high intensities of the unstable $N\approx Z$ parent nuclei required for these studies from $^{10}$C to $^{74}$Rb, which makes it well suited for precision decay studies.  Additionally, ISAC is in the unique position where world-class experimental equipment required to perform measurements of the required experimental quantities to high precision exist on site.

\subsection{Branching Ratio Measurements with the 8$\pi$ and\\GRIFFIN $\gamma$-ray Spectrometers}
Of the three experimental quantities that are required for Standard Model tests using superallowed Fermi $\beta$ decays, the measurement of the superallowed branching ratios to high-precision are arguably the most difficult.  For the studies performed at TRIUMF, we employ the method of precision $\gamma$-ray spectroscopy of excited nuclear states in the $\beta$-decay daughter nuclei.  In order to perform these measurements to high precision ($<0.1\%$), we rely on cases that have the dominant superallowed decay mode to the ground state in the daughter nucleus.  In these cases, we are able to characterize the total decay strength by observing $\gamma$-rays that result from the de-excitation of the few decays to excited states in the daughter nucleus; the precision of which is typically limited by the knowledge of the absolute $\gamma$-ray detection efficiency at a given energy.  Thus, by making relatively imprecise measurements of the strength of these individual weak transitions, we can subtract the result from unity to provide a high-precision determination of the dominant superallowed branch (cf. Refs.~\cite{Fin12,Lea08,Hyl06,Fin08,Dun13}).

At TRIUMF-ISAC we perform these measurements using the $8\pi$ $\gamma$-ray spectrometer, which consists of 20 Compton-suppressed, high-purity germanium (HPGe) detectors able to provide photo-peak efficiencies for single events of $\sim1\%$ at 1.3 MeV~\cite{Gar148pi}.  To detect coincident $\beta$ particles emitted in the decay of the parent nucleus, an array of 20 plastic scintillators (SCEPTAR) are used in the target chamber to dramatically improve the signal-to-background ratio of the $\gamma$-ray spectra.  For these measurements, a $\sim60$~keV continous ISAC beam is implanted on a movable tape system at the mutual centers of these two arrays where the decays occur.  The $8\pi$ has performed high-precision superallowed $\beta$-decay branching ratio measurements for $^{26m}$Al~\cite{Fin12}, $^{38m}$K~\cite{Lea08}, $^{62}$Ga~\cite{Hyl06,Fin08}, and $^{74}$Rb~\cite{Dun13}.

In 2014, after a decade of operation at ISAC, the $8\pi$ spectrometer was decommissioned to make way for the new high-efficiency Gamma-Ray Infrastructure For Fundamental Investigations of Nuclei (GRIFFIN) spectrometer.  GRIFFIN is an array of 16 large-volume ``clover-type" HPGe gamma-ray detectors, each of which contains 4 large germanium single crystals~\cite{Gar18}.  GRIFFIN is 17 times more efficient than the $8\pi$ for 1~MeV gamma rays and 37 times more efficient by 10~MeV.  The measurement principle with GRIFFIN is the same as with the $8\pi$, however with the significant increase in efficiency, new cases are now possible for measurement.  In fact, the first superallowed branching ratio measurement with GRIFFIN has been performed, and is currently under analysis ($^{22}$Mg).

\subsection{Half-Life Measurements with the GPS $4\pi$ Gas-Proportional Counter}
The superallowed $T_{1/2}$ measurements are performed using a $4\pi$ continuous-flow gas proportional counter that detects positrons from each decay as a function of time with nearly 100\% efficiency~\cite{Laf15}, and is located on the GPS beamline in the ISAC experimental hall.  The radioactive ions from ISAC are implanted under vacuum into a movable tape system that is made of aluminized Mylar. Following implantation, the tape is rapidly moved to position the sample in the center of the gas counter where the activity is typically measured for 25 half-lives of the respective radioactive species.  The tape cycles are individually tuned depending on the half-lives of the sample and any contaminant species that are delivered with the beam from ISAC.  The $T_{1/2}$ measurements with the gas counter system have been the most productive part of the experimental superallowed decay program at TRIUMF, due to the high-level of characterization we have achieved for the system as well as the relative simplicity of the experimental technique.  The most recent high-precision measurements in this campaign include $^{10}$C~\cite{Dun16}, $^{14}$O~\cite{Laf13}, $^{18}$Ne~\cite{Laf15}, $^{21}$Na~\cite{Fin17}, and $^{22}$Mg~\cite{Dun17}.

\subsection{Direct $Q_{EC}$-Value Measurements using the TITAN Penning Trap}
The $Q_{EC}$ value is extracted from the mass difference between the parent and daughter nuclear states in the superallowed decay mode, and can be measured to high precision using TRIUMF's Ion Traps for Atomic and Nuclear science (TITAN)~\cite{Dil06}.  The TITAN facility consists of four primary components: (i) A Radio-Frequency Quadrupole (RFQ) linear Paul trap~\cite{Smi06,Bru12}, (ii) a Multi-Reflection Time-of-Flight (MR-ToF) isobar separator~\cite{Jes15}, iii) an Electron-Beam Ion Trap (EBIT) for generating Highly Charged Ions (HCIs)~\cite{Sik05,Lap10} and performing decay spectroscopy~\cite{Lea17,Lea15,Len14}, and (iv) a 3.7~T, high-precision mass Measurement PEnning Trap (MPET)~\cite{Bro12}.  For experimental studies on RIBs with TITAN, ISAC delivers a continuous beam of ions at 20~keV which are injected into the TITAN-RFQ where they are trapped and cooled using a He buffer gas.  The resulting ion bunches are subsequently transported through the TITAN system with a kinetic energy of 2~keV to the Penning trap, where individual singly charged ions can be captured for study.  In MPET, the mass of a single ion can be determined to high precision by measuring its characteristic cyclotron frequency using the Time-of-Flight Ion-Cyclotron-Resonance (ToF-ICR) technique~\cite{Gra80,Kon95}.  For a direct $Q$-value determination, reference measurements are taken both before and after each run using the (typically stable) daughter nucleus.  For the determination of the resonance frequency ratios, only cycles with 1 detected ion/cycle are typically used in order to reduce effects on the measurement which may result from ion-ion interactions.  This is usually the largest systematic uncertainty in these measurements~\cite{Bro12}.  To date, TITAN has performed high-precision direct $Q_{EC}$-value measurements on the $T=1$ superallowed nuclei $^{10}$C~\cite{Kwi13}, $^{22}$Mg~\cite{Rei17}, and $^{74}$Rb~\cite{Ett11}. 


\section{Prospects for {\it ab-initio} $\delta_C$ Calculations}
\begin{figure}[t!]
\centering
\includegraphics[width=\linewidth]{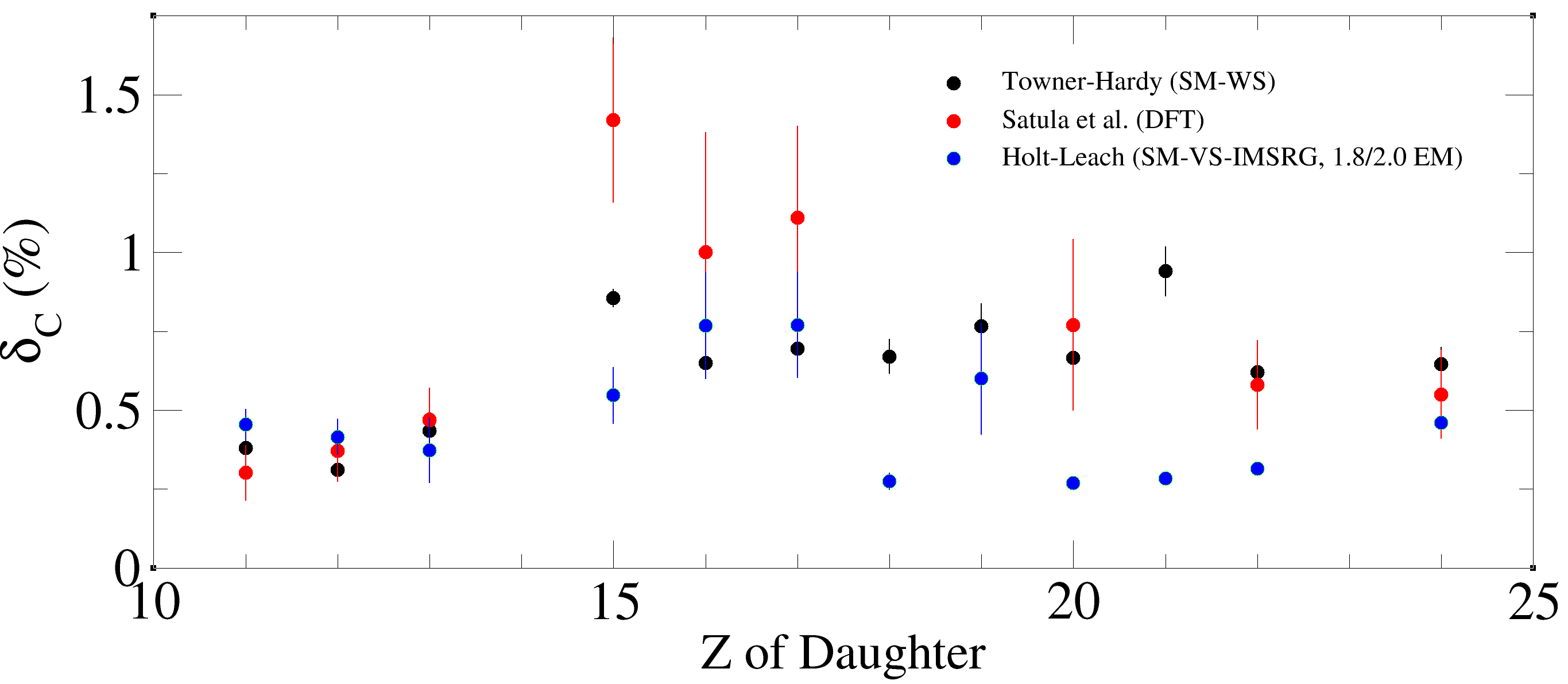}\\
\caption{(Color online) A comparison of our preliminary {\it ab-initio} theoretical $\delta_C$ calculations to the existing SM-WS~\cite{Tow08,Har15} and DFT~\cite{Sat16} methods for intermediate mass cases.}
\label{fig:delta_C}
\end{figure}
The current extraction of $G_V$ and $V_{ud}$ from the high-precision superallowed data uses the ISB corrections of Towner and Hardy, calculated within a semi-phenomenological shell model framework (TH)~\cite{Har15,Tow08}. This is largely due to the impressive efforts towards experimental testing~\cite{Bha08,Mel11,Par15} and guidance~\cite{Lea13a,Lea13b,Mol15} that their values have been exposed to. However, as experimental $ft$ values have become increasingly more precise, particularly in the last decade, model-space truncations~\cite{Tow10} and other theoretical approximations and deficiencies that exist in the TH formalism~\cite{Mil08} require renewed attention. Over the past two years, we have begun to explore the possibility of calculating ISB corrections to the experimental $\beta$-decay $ft$ values using {\it ab-initio} theoretical techniques, both as a consistency check for the existing TH formalism~\cite{Tow08} and to gain additional insights from the direct connection to underlying two- (NN) and three-nucleon (3N) forces. 

Traditionally most nuclei relevant for superallowed studies have been beyond the scope of {\it ab-inito} methods, which were either restricted to $A \approx 12$~\cite{Nav97} or to regions near closed shells~\cite{Hage14RPP}. These constraints have been largely overcome in the past few years years with the development of many-body techniques tailored towards medium-mass open-shell systems, such as the valence-space in-medium similarity renormalization group (VS-IMSRG)~\cite{Bogn14SM,Stro16TNO,Stro17ENO} discussed here. Based on NN and 3N forces forces from chiral effective field theory (EFT)~\cite{Epel09RMP,Mach11PR}, {\it ab-initio} methods are beginning to approach levels of accuracy comparable of phenomenological models in the $sd$ and $pf$ shells~\cite{Stro16TNO,Lea16,Rei17}.

There are a number of advantages to the {\it ab-initio} approach, primarily the fact that the Fermi matrix element $|M_F|$ can be calculated directly to extract $\delta_C$, without the need for the separation of terms used in the TH formalism~\cite{Mil08,Tow08}. Additionally, these methods can potentially quantify systematic uncertainties or possible shifts in the $\delta_C$ central values, which still remain elusive due to the extreme complexity of the phenomenological approach to the nuclear shell model. While this work is still in the preliminary stages, in Fig.~\ref{fig:delta_C} we show a first comparison of VS-IMSRG calculations of $\delta_C$ to the TH Shell-Model Woods-Saxon (SM-WS) values from Ref.~\cite{Har15} and the Density Functional Theory (DFT) approach of Ref.~\cite{Sat16}. The input Hamiltonian is the 1.8/2.0 (EM) NN+3N interaction developed in Refs.~\cite{Hebe11fits,Simo17SatFinNuc}, which has been shown to well reproduce ground-state energies from the $p$-shell to the tin region~\cite{Morr18Sn}. In addition we use effective valence-space operators consistently transformed with the Hamilotonian~\cite{Parz17Trans}. While we find general agreement with the results of other methods, theoretical error estimation remains the primary challenge. Here the error bars quantify one such uncertainty arising from the arbitrary option of normal ordering operators in the VS-IMSRG approach with respect to either the parent or daughter nucleus~\cite{Stro17ENO}. While largely expected to be negligible, this choice is clearly important for particular nuclei. This only emphasizes the need for a careful and systematic exploration of theoretical uncertainties, including those from truncations in the chiral EFT expansion~\cite{Car16}, many-body approximations, and basis-space convergence.
 
As these theoretical techniques continue to evolve, they must be exposed to increasingly stringent experimental tests before they can be reliably confronted with the superallowed data to extract $V_{ud}$.  In particular, a reproduction of the coefficients of the isobaric-multiplet-mass-equation (IMME) for the respective superallowed systems are critical to providing confidence in the accuracy of the calculated ISB corrections. The coefficients of the IMME are sensitive to the subtle relative differences in binding energies of the isobaric triplet, and have been used to guide and adjust the TH superallowed $\delta_C$ calculations in the past~\cite{Tow08}. Investigations along these lines are currently in progress.

\section{Conclusions}
The high-precision experimental superallowed $0^+\rightarrow0^+$ $\beta$ decay program at TRIUMF has been one of the flagship scientific efforts at ISAC over the past 15 years.  This program has been instrumental in reducing the overall uncertainty on the up-down element in the CKM matrix to the currently reported level of 0.021\%~\cite{PDG16}, as well as providing stringent limits on possible scalar contributions to the vector part of the weak interaction~\cite{Dun16}.  As ISAC continues to improve its capability to deliver high-intensity, pure samples of $N\approx Z$ nuclei, and with ongoing improvements to the experimental equipment, the future of this program continues to be bright. From a nuclear structure theory standpoint, {\it ab-initio} approaches to calculating $\delta_C$, which do not require the separation of terms employed in the TH formalism, are currently under development. These efforts are still in their infancy, but currently show a promising level of agreement in the $sd$ and lower $pf$ shells. These calculations will help to provide a consistency check to the existing ISB corrections to the high-precision superallowed data for weak-interaction testing.


\Acknowledgements
This work was supported in part by the U.S. Department of Energy Office of Science under contract DE-SC0017649, and the Natural Science and Engineering Research Council of Canada (NSERC).  TRIUMF receives federal funding via a contribution agreement with the National Research Council of Canada (NRC).  We would like to thank Gordon Ball, Jens Dilling, Michelle Dunlop, Adam Garnsworthy, G.F. Grinyer, Ania Kwiatkowski, Alex Laffoley, Jens Lassen, Erich Leistenschneider, Hamish Leslie, Petr Navratil, Pascal Reiter, Achim Schwenk, Ragnar Stroberg, and Carl Svensson for discussions, slides or content that were presented during the conference or included in this proceeding.

\footnotesize
\bibliography{references}

\end{document}

%% file: econfmacros.tex
\def\beq{\begin{equation}}
\def\eeq#1{\label{#1}\end{equation}}
\def\eeqn{\end{equation}}

\def\beqa{\begin{eqnarray}}
\def\eeqa#1{\label{#1}\end{eqnarray}}
\def\eeqan{\end{eqnarray}}

\let\bar=\overbar

\def\Dslash{\not{\hbox{\kern-4pt $D$}}}
\def\dslash{\not{\hbox{\kern-2pt $\del$}}}

\def\msb{{\bar{\ssstyle M \kern -1pt S}}}

